\documentclass[fleqn,twoside]{article}
\usepackage[headings]{espcrc2}
\readRCS
$Id: espcrc2.tex,v 1.2 2004/02/24 11:22:11 spepping Exp $
\usepackage{graphicx}
\usepackage[figuresright]{rotating}

\newcommand{\AmS}{{\protect\the\textfont2
  A\kern-.1667em\lower.5ex\hbox{M}\kern-.125emS}}
\title{Formation of neutron-rich and superheavy elements in 
       astrophysical objects}

\author{S.K. Patra\address[IOP]{Institute of Physics, 
       Sachivalaya Marg, Bhubaneswar-751 005, India }%
        and
        R.N. Panda\address[REC]{Department of Physics, Raajdhani Engineering 
        College, Mancheswar, Bhubaneswar-751 017, India.}}%
\runauthor{S.K. Patra and R.N. Panda}

\begin{document}

\begin{abstract}

We calculate the reaction and the fusion cross-sections of neutron-rich
heavy nuclei taking light exotic isotopes as projectiles. Results of
 neutron-rich
Pb and U isotopes are demonstrated as the representative targets and
He, B as the projectiles. The Gluaber Model and the Coupled Channel Formalism
are used
to evaluate the reaction and the fusion cross-sections for the cases considered.
Based on the analysis of these cross-sections,
we predict the formation of heavy, superheavy and super-superheavy
elements through rapid neutron/light nuclei capture r-process of
the nucleosynthesis in astrophysical objects.
\vspace{1pc}
\end{abstract}
\maketitle

Formation of superheavy elements (SHE) in the laboratory is one of
the most challenging
problem in Nuclear Physics. So far the synthesis  of  Z=118
element has been possible \cite{ogan06}. Efforts are on to synthesise still
 heavier elements in various laboratories all over the world. It is certain that 
if an element is created through human efforts then definitely it must be
present naturally somewhere in the Universe. Thus the mode of formation of superheavy or
super-superheavy element in astrophysical object is a fundamental
question in the field of Nuclear Astrophysics. In this context, it is
mandatory that the superheavy element with Z=118 and higher atomic numbers
are present in the object like relativistic jets of $\gamma-rays$ bursts (GRBs)
or supernovae jets near the nascent neutron star. It has been
reported in Ref.\cite{moller92}, and the stability of the most stable superheavy elements could be as high as  $10^9$ years in some of the calculations \cite{nilsson68}.

Study of unstable nuclei with radioactive ion beam (RIB) facilities
has opened an exciting channel to look upto  some of the crucial issues in
the context of both nuclear structure and astrophysics
\cite{oza01}. Unstable nuclei play an influential, and
in some cases dominant role, in  phenomena of the cosmos such as
Gamma Ray Bursts.

The direct study of stellar properties in ground-based
laboratories has become feasible, due to the availability of
RIBs; for example the study of $^{18}$Ne induced neutron pick-up
reaction could reveal information about the exotic
$^{15}$O+$^{19}$Ne reaction occuring in the CNO cycle in 
stars. Study of the structure and the reactions of not only unstable
light exotic but also of the superheavy and the super-superheavy
nuclei is therefore required to improve our understanding of
the astrophysical origin of atomic nuclei and the evolution of stars and their
 death.

In a recent study, Satpathy et al.
\cite{patra08} claimed the neutron-rich U and
Th-isotopes are
thermally fissile and could release orders of magnitude more
energy than $^{235}$U in a new mode of fission decay called
{\it multi-fragmentation fission}, which happend frequently in
astrophysical objects. The main objective of the present letter is
to study the reaction ($\sigma_r$) and fusion ($\sigma_f$)
cross-sections of neutron-rich U and some other interesting exotic
isotopes, which are related to the formation of neutron-rich,
SHE and super-SHE elements in the Universe.

The value of $\sigma_r$ is calculated by
using the most recently developed effective field theory motivated
relativistic mean field (E-RMF) nuclear densities \cite{tang96},
in conjunction with
the Glauber model. However, $\sigma_f$ is estimated in the non-relativistic
coupled channel calculation. From the calculated reaction and
fusion cross-sections, we look for the
formation path of neutron-rich, SHE and super-SHE
nuclei in the cosmos.
The theoretical formalism to calculate
the nuclear reaction cross-section using Glauber approach has been given
by R. J. Glauber \cite{gla59}. The standard Glauber form for the reaction
cross-section at high energies, is expressed \cite{gla59} as:
\begin{equation}
\sigma _{r}=2\pi\int\limits_{0}^{\infty }b[1-T(b)]db \;,
\end{equation}
where $T(b)$, the transparency function, is the probability that at
an impact parameter $b$ the projectile passes through the target
without interaction. This function $T(b)$ is calculated in the
overlap region between the projectile and target where the
interactions are assumed to result from single nucleon-nucleon
collision and is given by
\begin{equation}
T(b)=\exp \left[ -\sum\limits_{i,j}\overline{\sigma }_{ij}\int d%
\vec{s}\overline{\rho }_{ti}\left( s\right) \overline{\rho }%
_{pj}\left( \left| \vec{b}-\vec{s}\right| s\right)
\right] \;.
\end{equation}
Here, the summation indices $i$, $j$ run over proton and neutron numbers and
subscript $p$ and $t$ refers to projectile and target respectively.

The original Glauber model is designed for high energy projectile,
like relativistic proton reactions. It fails to describe 
the collisions induced at relatively low energies. In this case, the
straight-line trajectory is modified because of the presence of the
Coulomb field of the target and projectile. In such cases the present 
version of Glauber model is modified in order to take care of finite range 
effects\cite{shuk03}  in  the profile function and the Coulomb 
modified trajectories.  Thus for 
finite range approximations, the transparency function is given by

\begin{eqnarray}
T(b)&=&\exp \left[ -\int\nolimits_{P}\int\nolimits_{T}\sum\limits_{i,j}\left[
\Gamma _{ij}\left( \vec{b}-\vec{s}+\vec{t}%
\right ) \right ] \right . \nonumber \\
& & 
\left .\left .
\overline{\rho }_{Pi}
\left( \vec{t}\right)
\overline{\rho }_{Tj}\left( \vec{s}\right) d\vec{s}d%
\vec{t}\right .\right]\;.
\end{eqnarray}

Here the profile function $\Gamma _{ij}$ is given by
\begin{equation}
\Gamma _{ij}(b_{eff})=\frac{1-i\alpha }{2\pi \beta _{NN}^{2}}\sigma
_{ij}\exp \left( -\frac{b_{eff}^{2}}{2\beta _{NN}^{2}}\right)\;,
\end{equation}
where $b_{eff}=\left| \vec{b}-\vec{s}+\vec{t%
}\right| $,
$\vec{b}$ is the impact parameter and $\vec{s%
}$ and $\vec{t}$ are just the dummy variables for integration over
the $z$-integrated target and projectile densities. 
The values of the parameters, ${\bar{\sigma}}_{ij}$, $\alpha$ and
$\beta_{NN}$ are taken from Ref. \cite{far01}.
The detailed formalism is available in Ref.\cite{patra07}.

The E-RMF density with G2 parameter set \cite{tang96,patra01a} is used as
input for the evaluation of
$\sigma_r$.  For the details of the calculation of ground state properties
of finite nuclei and the procedure of estimation of nuclear cross-section,
we refer the reader to Refs. \cite{patra07,patra01a,ibrahim03}.

To compute the fusion cross-section $\sigma_f$ we follow the coupled-channel
calculations including all orders of coupling. This is done in a non-relativistic
framework. The computer code CCFULL as developed in Ref. \cite{hagino99} is
used. The fusion cross-section is given by the formula \cite{hagino99}:
\begin{equation}
\sigma_f(E)=\sum_J\sigma_J(E)=\frac{\pi}{k_0^2}\sum_J(2J+1)P_J(E),
\end{equation}
with $P_J(E)$ is the inclusive penetrability and the other symbols have
the standard meaning as defined in \cite{hagino99}.

It was shown
in our earlier papers that the densities taken from relativistic mean field
formalism, and used in the frame-work of Glauber model \cite{gla59,ibrahim03} to
evaluate the differential and total reaction cross-section is quite successful
for light systems \cite{patra07}. Now we extend the model to
calculate the total reaction cross-section considering light exotic nuclei
as projectile and heavy neutron-rich isotopes as target. Here,
we calculate as the representative cases for the reaction cross-section
of neutron-rich Pb and U isotopes taking exotic He and B nuclei as
incident projectile.

In Fig. 1 the reaction cross-section $\sigma_r$ for
$^{4}$He+$^{208,228,248,278}$Pb,  $^{10,15,17,20}$B+$^{208}$Pb,
$^{4}$He+$^{235,250,270,290}$U and $^{10,15,17,20}$B+$^{235}$U are presented.
From the calculated results, the increase in $\sigma_r$ is quite substantial
with the target mass. The same observation is also applicable, while increasing
the mass of the projectile (keeping the target mass constant). In any of these
cases, the reaction cross-section becomes favourable with either increase
of projectile mass or the mass of the target or both. The enhancement
can be understood by the simple classical expression of the 
cross-section $\pi R^2$ ($R$=radius of the nucleus) where the increase is 
due to the larger
size of the nucleus. This implies the
probability of formation of heavier masses in the reaction process with
heavier isotope of the projectile as well as target. In Ref. \cite{de01},
within the formalism of a Thomas-Fermi
model, calculations are presented for nuclei beyond the nuclear drip-line at
zero temperature. This is possible because of the presence of an external
neutron gas which may be envisaged in the astrophysical scenario and is 
the situation of the present discussion for accreting cosmological objects.

\begin{figure}[ht]
\begin{center}
\includegraphics[width=1.0\columnwidth]{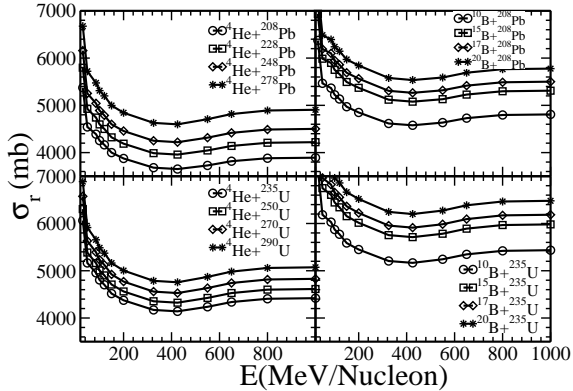}
\caption{{\it The nuclear reaction cross-sections taking He and B isotopes
as projectile with different isotopes of Pb and U.}}
\end{center}
\end{figure}
\begin{figure}[ht]
\begin{center}
\includegraphics[width=1.0\columnwidth]{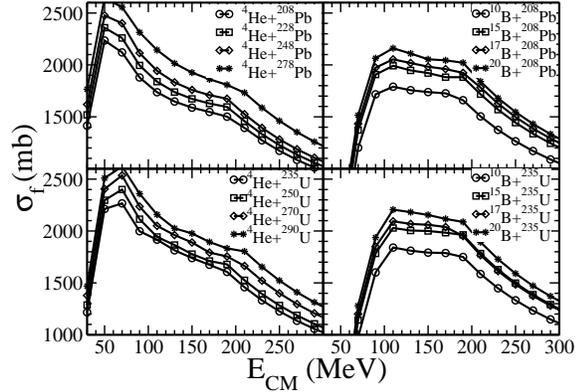}
\caption{{\it The nuclear fusion cross-sections taking He and B isotopes
as projectile with different isotopes of Pb and U.}}
\end{center}
\end{figure}
\begin{figure}[ht]
\begin{center}
\includegraphics[width=1.0\columnwidth]{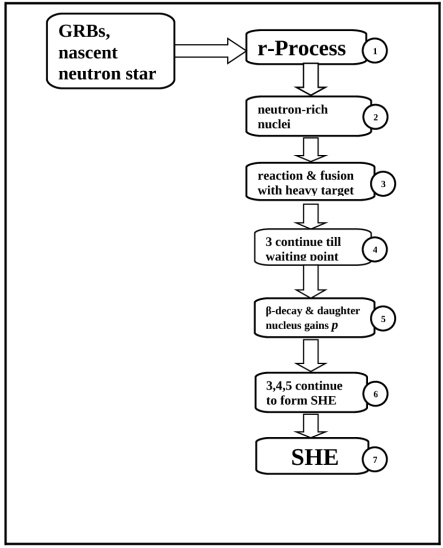}
\caption{{\it The schematic diagram for the formation of superheavy
element (SHE)
in the astrophysical object such as relativistic jets of GRBs or supernovae jets near the nascent neutron star. The
production of SHE is possible through reaction and fusion processes at
a favourable energy condition in the cosmos.}}
\end{center}
\end{figure}
\begin{figure}[ht]
\begin{center}
\includegraphics[width=1.0\columnwidth]{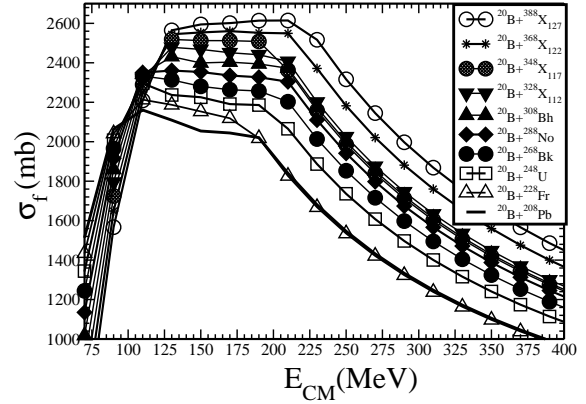}
\caption{{\it A representative path for the formation of $^{408}X_{132}$
superheavy element through $^{20}$B capture process. The fusion
cross-sections $\sigma_f$ for various daughter nuclei with $^{20}$B is shown.
}}
\end{center}
\end{figure}

In Fig. 2 the fusion cross-section $\sigma_f$ for various neutron-rich
light nuclei with heavier drip-line isotopes, like
$^{4}$He+$^{208,228,248,278}$Pb,  $^{10,15,17,20}$B+$^{208}$Pb,
$^{4}$He+$^{235,250,270,290}$U and $^{10,15,17,20}$B+$^{235}$U are shown.
Similar to the reaction cross-section, the increase in $\sigma_f$ is quite
clear with the increase of target, projectile or both the
masses.  This implies the probability of
creation of heavier masses with the increase of mass number of the projectile as
well as target and making the way for the evolution of neutron-rich
heavy nuclei much beyond the drip-line \cite{de01} due to the presence of
the  external neutron gas or highly neutron-rich light as well as heavy nuclei
generates in the astrophysical objects,in the relativistic jets of GRBs or
 supernovae jets near the nascent neutron star.
Analysis of figures 1 and 2 shows that, the magnitude of $\sigma_r$ and $\sigma_f$
are optimum at $\sim 30$ to 200 MeV of the incident projectile energy. Beyond
this range, the value of $\sigma_r$ and $\sigma_f$ decreases drastically.
Both the cross-sections indicate the suitability of the incident projectile
energy for a favourable condition of the formation of the fused elements
in the astrophysical system. Thus, the chance of the formation of
heavier element is maximum, if a suitable
energy range is created, which may be a source in  the relativistic 
jets of GRBs or  supernovae jets near the nascent neutron star \cite{mazz06,thom04}.
The high energy environment in such cosmological objects is
 because of the
supernova shock \cite{laga83} and it is quite common in the nascent
 neutron star or 
relativistic jest of GRBs \cite{mazz06,thom04}. In these objects a highly 
neutron-rich and high temperature 
scenario is made possible and  which may be a probable platform for such reactions.

In this context, it is worth citing the  following example: A 
neutron star is borned when a star of mass
$\sim 20$ $M_{\odot}$ undergoes its core collapses after hyper-energetic explosions
of Gamma ray bursts. A star with initially $\sim 20$ $M_{\odot}$ would develop 
carbon-oxygen core of  $\sim 3.3$ $M_{\odot}$. It left behind a neutron star of
$\sim 1.4$ $M_{\odot}$, $\sim 1.3$ $M_{\odot}$ of oxygen and $\sim 0.6$ 
$M_{\odot}$ of heavier 
elements, Si and Fe group,  which  could be ejected in the supernova.
Such a collapse gives rise to an explosion of kinetic energy 
K.E. $\sim 10^{51}$ ergs ($\sim 6.25\times 10^{56} MeV$)
\cite{mazz06,thom04}.

Young neutron stars have a fluid surface, a solid, crystalline crust and a
fluid interior. The fluid regions of the star adjust themselves to its 
rotation which remaining always axi-symmetric. The radiated power comes directly 
from the rotational energy of the neutron star.  The entropy in mass elements 
exhibiting the neutron star at later times will be larger than  
the earlier. This is because, most of the heating occurs 
near the surface of the neutron star. Slowly with time the radius of 
the neutron star shrinks from $~100 Km$ to $10 Km$ \cite{meyer94}. The decrease in the initial 
radius start from which the mass elements begin increasing the heat rate 
\cite{mazz06,thom04}.

It is worth mentioning of the burning process of H, He, Li, .... in the
accreting astrophysical system. To maintain {\it hydrostatic
equalibrium} \cite{pearson86}, this continues upto formation
of Iron. When this stage is reached, depending on its
mass, the astrophysical object undergoes various phenomena like
supernovae explosion, X-rays burst, GRBs, formation of neutron star, black hole, red giant or white dwarf  etc.
In some cases, it becomes highly neutron-rich environment (novae, supernovae or
X-rays burst or neutron star) and is favourable for rn-process, which
continue upto certain A or Z number. Slowly, this rn-capture process
becomes less favourable and fusion of light nuclei (like He, Li, Be,...)
become more important.
	In the mean time, the neutron-rich light element fused with
these heavy nuclei and more heavier isotopes with a little increase of
proton number is generated in the process; for example, $^{4}$He+$^{208}$Pb
gives $^{212}$Po.  Again  $^{212}$Po reacts with $^{4}$He to
form $^{216}$Rn. A schematic diagram for the process of SHE
formation is shown in Figure 3. From the figure, it can be understood how
this phenomenon goes on  to create  much heavier
isotopes.
Similarly other processes also continue to go on as shown in figures 1, 2 and
3, such as
$^{20}$B+$^{235}$U$\rightarrow$ $^{255}$Bk,
$^{20}$B+$^{255}$Bk $\rightarrow$$^{275}$No, ..... and so on.

A representative example is depicted in Figure 4. As mentioned earlier, after
the supernovae explosion, in the rn-process, heavy normal/exotic nuclei
including the ultra-neutron-rich light isotopes are formed. Exotic
nuclei like $^{6}$He, $^{11}$Li, $^{14}$Be, $^{20}$B, normal actinides
(e.g. $^{208}$Pb, $^{235}$U etc.)  and neutron-rich
drip-line isotopes, similar to $^{278}$Pb etc. are
generated. Thereafter, fusion process of the light isotopes with heavier
nuclei becomes important. The increase of fusion cross-sections as shown
in Fig. 4 confirmed the possibility of the formation of ultra-heavy isotopes
as well as superheavy elements both with lower and higher atomic masses.
The demonstration of a path for the formation
of $^{408}X_{132}$ (A=408, Z=132, N=276) through complete fusion process is
given below (whose cross-sections are shown in Fig. 4):
$^{20}$B+$^{208}$Pb $\rightarrow$ $^{228}$Fr,
$^{20}$B+$^{228}$Fr $\rightarrow$ $^{248}$U,
$^{20}$B+$^{248}$U $\rightarrow$ $^{268}$Bk,
$^{20}$B+$^{268}$Bk $\rightarrow$ $^{288}$No,
$^{20}$B+$^{288}$No $\rightarrow$ $^{308}$Bh,
$^{20}$B+$^{308}$Bh $\rightarrow$ $^{328}X_{112}$,
$^{20}$B+$^{328}X_{112}$ $\rightarrow$ $^{348}X_{117}$,
$^{20}$B+$^{348}X_{117}$ $\rightarrow$ $^{368}X_{122}$,
$^{20}$B+$^{368}X_{122}$ $\rightarrow$ $^{388}X_{127}$,
$^{20}$B+$^{388}X_{127}$ $\rightarrow$ $^{408}X_{132}$ and so on.

Thus, each time the proton number Z increases by 5 units the mass number
A goes up by 20 units 
in the case of $^{20}$B capture. Slowly, it creates a highly neutron-rich
heavy isotope, which is enabled to capture any more neutron $n$ or
neutron-rich nucleus. This is
termed as {\bf{\it waiting point}}. Here, the neutron-rich heavy element 
emits a $\beta^--$ particle, and the daughter nucleus gains
a positive charge by converting a neutron ($n$) to a proton ($p$). Due to
this enhancement
in $Z$, the product (daughter nucleus) captures few more
$n$ or neutron-rich light nuclei by fusion process  till it reaches the new waiting
point. At this point, the nucleus gains another proton $p$, by
emitting $\beta^--$particle.
This process continues and SHE or super-SHE
are formed in the cosmoligical object.
In this context, it is worth mentioning that, the
dominant mode of decays are $\beta^-$ and
spontaneous fission for large N and large Z nuclei, respectively.
In the $\beta^--$
decay, the daughter nucleus gains a proton, whereas for large
N, the spontaneous fission reduces considerably due to excess
number of neutrons \cite{patra08} and the neutron-rich isotope
becomes fission stable as the height of the fission barrier decreases
and the width increases, thereby making the nucleus more stable against fission
\cite{patra08}.  It is interesting to mention here that,
recently it has been  reported by A. Marinov et al. \cite{marinov08},
that the evidence of a superheavy
isotope with $Z=122$ or 124 and a mass number A=292; has been found in natural
Th using inductively coupled plasma-sector field mass spectrometry. The
estimated half-life of this isotope is $t_{1/2}\ge 10^8$ years, comparable
with the theoretical predictions \cite{nilsson68}.

In summary, we estimated the reaction and fusion cross-section of various
combination of light and heavy isotopes. We extended the calculations to
exotic systems taking into consideration the possibility of availing the
rn-process and the exotic nuclei capture processes
in astrophysical objects. The enhanced cross-sections with increase of mass
number for both the projectile and target made it possible for the
formation of the heavier neutron-rich nuclei way beyond the normal
drip-lines predicted by the mass models. By the neutron or heavy ion
(light neutron-rich nuclei)  capture process the daughter nucleus becomes a
superheavy element which may be available somewhere in the
Universe in super-natural condition and can be possible to be synthesised in
laboratories. Here the stability of the neutron-rich SHE or super-SHE against
spontaneous fission arises due to widening of the fission barrier because
of the excess number of neutrons.

\noindent {\bf Acknowledgement}

Discussions with Professors K. Langanke, L. Satpathy and C.R. Praharaj
are gratefully acknowledged. We are thankful to Prof. A. Abbas for a
careful reading of the manuscript.
This work has been supported in part by Council of Scientific $\&$ Industrial
Research (No. 03(1060)06/EMR-II) as well as the project No. SR/S2/HEP-16/2005,
Department of Science and Technology, Govt. of India.

\end{document}